\documentclass[fleqn,10pt]{SelfArx} 

\definecolor{color1}{RGB}{0,0,90} 
\definecolor{color2}{RGB}{0,20,20} 

\usepackage{hyperref} 
\hypersetup{hidelinks,colorlinks,breaklinks=true,urlcolor=color2,citecolor=color1,linkcolor=color1,bookmarksopen=false,pdftitle={Title},pdfauthor={Author},urlcolor=blue}

\usepackage{graphicx}
\usepackage[square]{natbib}
\usepackage{amsmath}
\usepackage{multirow}
\usepackage{subcaption}
\usepackage{pdflscape}
\usepackage{epstopdf}
\usepackage[final]{pdfpages}

\usepackage{graphicx}
\usepackage{natbib}
\usepackage{amsmath}
\usepackage{amsfonts}
\usepackage{multirow}
\usepackage{tikz}
\usetikzlibrary{calc}

\usepackage{mathtools,amssymb,lipsum}

\usepackage{cuted}
\usepackage{boondox-calo}

\usepackage{calligra}

\newcommand{\bs}[1]{\boldsymbol{#1}}

\newcommand{\n}[1]{\mathrm{#1}}

\newcommand{\ud}{\mathrm{d}}

\JournalInfo{Published in Journal of Magnetism and Magnetic Materials, Vol. 632, 173519, 2025} 
\Archive{\href{https://doi.org/10.1016/j.jmmm.2025.173519}{DOI: 10.1016/j.jmmm.2025.173519}} 

\PaperTitle{The magnetic scalar potential and demagnetization vector for a cylinder tile} 

\Authors{Rasmus Bjørk} 
\affiliation{\textit{Department of Energy Conversion and Storage, Technical University of Denmark - DTU, DK-2800 Kgs. Lyngby, Denmark}} 
\affiliation{\textbf{Corresponding author}: rabj@dtu.dk} 

\newcommand{\keywordname}{Keywords} 

\Abstract{A closed-form solution for the magnetic scalar potential generated by a uniformly magnetized cylindrical slice and a full cylinder is determined by solving Poisson's equation analytically. The solution is given in terms of elliptic integrals of the first, second and third kind. We show that the magnetic scalar potential can be written as the dot product of a demagnetization vector, containing all the geometric information of the generating cylinder, and the magnetization. We validate the derived analytical expressions for the magnetic scalar potential by comparing with a finite element simulation and show that these agree perfectly for both the cylindrical slice and the full cylinder.}

\begin{document}

\flushbottom 

\maketitle 


\thispagestyle{empty} 
\section{Introduction}
Magnetic fields are fundamental in nature, as well as essential components in a plethora of technologies including electric motors and generators to fusion reactors. Permanent magnets are increasingly used to generate magnetic fields up to 2 T for these  technologies. 

The magnet field generated by a permanent magnet depends on the shape of the magnet. If the magnet is uniformly magnetized the geometrical dependence can be captured in the demagnetization tensor, as the dot product of this and the magnetization gives the magnetic field at a specific point. The demagnetization tensor can for some objects be calculated analytically such as for example for a prism \cite{Smith_2010}, cylinder \cite{Nielsen_2020,Slanovc_2022}, tetrahedron \cite{Nielsen_2019}, ellipsoid \cite{osborn_1945,Tejedor_1995} and hollow sphere \cite{PratCamps_2016}. This is very useful as it allows the magnetic field to be computed with arbitrarily high precision, which can be required for applications in geophysics \cite{maus2009emag2,mclay2010interpolation}, magnetohydrodynamics \cite{mackay2006divergence}, particle physics experiments \cite{bernauer2016measurement}, navigation \cite{le20123, Solin_2018} and astronomy \cite{sun2011new}.

However, as the magnetic field is a vector field, the analytical expressions for the $3\times{}3$ demagnetization tensor can become very complex. Additionally, in the growing field of physics-informed machine learning, there are significant advantages to working with scalar potentials, as the fields derived from these are conservative by construction, and only a single loss function is required to learn the potential \cite{Solin_2018}. This prevents tedious balancing of the multiple losses \cite{Bischof_2025} arising from the different field components and other constraints, e.g., energy conservation.

If there are no free currents present in the magnetic system studied, as is the case for permanent magnets, instead of considering the magnetic field, one can consider the magnetic scalar potential, which is defined such that the negative gradient of this scalar potential is the magnetic field. It should be mathematically possible to derive analytical expressions for this scalar potential, which is a single-valued function, for various shapes of permanent magnets if these are uniformly magnetized, yet this has not been done in literature until now except for simple geometries. Recently the analytical expression for the magnetic scalar potential for a uniformly magnetized prism was derived \cite{James_2025}, and it was shown that the magnetic scalar potential can be written as the product of the demagnetization vector, containing the geometric information, and the magnetization. Additionally, the magnetic scalar potential is also known for a magnetic dipole \cite{James_2025} as well as for a uniformly magnetized sphere \cite{James_2025}. The gravitional potential for a prism, which can be expressed by an equation very similar to that present in magnetostatics, is also known \cite{Chappell_2013}. In this work, we derive the magnetic scalar potential for a uniformly magnetized cylindrical slice and a full cylinder and compare these to a numerical model to validate the derived analytical expressions. 

\section{Theory}
We assume that there are no free currents in the magnetic system that we consider, as is the case for permanent magnets. This means that the magnetic field, $\mathbf{H}$, is irrotational. We consider a single uniformly magnetized body termed $\Omega$. In the whole space, the magnetic field from this object can be expressed as the gradient of a scalar field, termed the magnetic scalar potential, $\Psi_M$, as
\begin{equation}
\mathbf{H}  = -{\nabla}\Psi_M\label{eq:PhiM}.
\end{equation}

To express the magnetic scalar potential as a function of the magnetization, $\mathbf{M}$, which is what generates the field, we employ the relation between the magnetic field, $\mathbf{H}$, the magnetic flux density, $\mathbf{B}$, and the magnetization, $\mathbf{M}$, which is 
\begin{equation} \mathbf{B} = \mu_0(\mathbf{H} +\mathbf{M} )\label{eq:Hdef}\end{equation}
where $\mu{}_{0}$ is the vacuum permeability. 

Inserting this relation in Eq. \eqref{eq:PhiM}, taking the divergence and using Gauss's law for magnetism, ${\nabla}{}\cdot{}\mathbf{B}=0$, we obtain the Poisson equation for the magnetic scalar potential
\begin{equation}
-{\nabla}\cdot{}(\mu{}_{0}\nabla\Psi_M)=-\nabla\cdot{}(\mu{}_{0}\mathbf{M} )~.\label{Eq.Poisson01}
\end{equation}

Noting that the magnetization is assumed to be nonzero only within the region $\Omega$, the above equation must satisfy the following condition on the boundary of $\Omega$, termed the surface $S'$,
\begin{equation}
\mathbf{\hat{n}}\cdot[-\nabla\Psi_M]_{S'}=\mathbf{\hat{n}}\cdot \mathbf{M}~. \label{Eq.Poisson02}
\end{equation}
Here the outward normal on the surface $S'$ is denoted by $\mathbf{\hat{n}}$ and with $[-\nabla\Psi_M]_{S'}$ the jump of the vector field $\mathbf{H}=-\nabla\Psi_M$ across $S'$. Comparing to electrostatics one can identify the volume and surface \emph{equivalent magnetic charge} densities $\rho_M=-\nabla\cdot\mathbf{M}$ and $\sigma_M=\mathbf{\hat{n}}\cdot \mathbf{M}$, respectively.

We here consider an object with homogeneous magnetization throughout, i.e.  $\mathbf{M}$ is constant in $\Omega$. This means that $\rho_M=0$ and the magnetic scalar potential is produced only by the surface charge density $\sigma_M=\mathbf{\hat{n}}\cdot \mathbf{M}$. The solution to Poisson's equation is then given by \cite{Jackson}:
\begin{equation}
    \Psi_M(\bs{r})=\frac{1}{4\pi} \oint_{S'}\frac{\mathbf{\hat{n}}(\mathbf{r}')\cdot\mathbf{M}}{\|\mathbf{r}-\mathbf{r}'\|}\ud S'~.\label{eq:PhiMsolFlat}
\end{equation}
We note that the above equation has two sets of coordinates. The coordinates marked with a $'$ are the coordinates of the face that creates the magnetic field, whereas the non-marked coordinates are to the point at which the field is evaluated. Once the magnetic scalar potential has been determined, the magnetic field can be determined from Eq. \eqref{eq:PhiM}. In the above integral we take the constant of integration to be zero, corresponding to a gauge of zero, equivalent to $\Psi_M\xrightarrow[\mathbf{r} \to \infty]{} 0$.

As the magnetization is constant across a surface, and thus does not depend on $\mathbf{r}'$, it can be moved outside the integral in Eq. \eqref{eq:PhiMsolFlat} and we get
\begin{equation}
    \Psi_M(\bs{r})=\frac{1}{4\pi} \oint_{S'}\frac{\mathbf{\hat{n}}(\mathbf{r}')}{\|\mathbf{r}-\mathbf{r}'\|}\ud S'\cdot\mathbf{M}~,\label{eq:PhiMsolFlat2}
\end{equation}
where it is understood that the integral is of each component of the normal vector. As was noted in Ref. \cite{James_2025}, this is the magnetic scalar potential in vector notation, which can be reformulated as
\begin{equation}
\Psi_{M}(\bs{r})=\mathbb{N}_\Psi(\bs{r})\cdot{}\mathbf{M}~, \label{Eq.final}
\end{equation}
where the demagnetization vector for the magnetic scalar potential was introduced as $\mathbb{N}_\Psi(\bs{r})$. This contains all the geometric information necessary for calculating the magnetic scalar potential and subsequently the magnetic field and will be derived in the following for the cylindrical slice and the full cylinder.

We note that Refs. \cite{Nielsen_2020,Slanovc_2022} analytically calculate the magnetic field, and not the magnetic scalar potential, for the cylindrical geometry studied here. While the magnetic field is the gradient of the magnetic scalar potential, in analytically calculating the field one can interchange the gradient operator and the integral operator in Eq. \eqref{eq:PhiMsolFlat2} as these are with respect to the coordinates of the evaluation point and of the magnetized source, respectively \cite{Nielsen_2020}. This means that Refs. \cite{Nielsen_2020,Slanovc_2022} does not analytically calculate the magnetic scalar potential, even though they determine the magnetic field, which clearly distinguish this work from Refs. \cite{Nielsen_2020,Slanovc_2022}.

\section{The magnetic scalar potential of a cylinder slice}\label{app:face}
\begin{figure}[ht]
  \centering
  \includegraphics[width=1.0\linewidth]{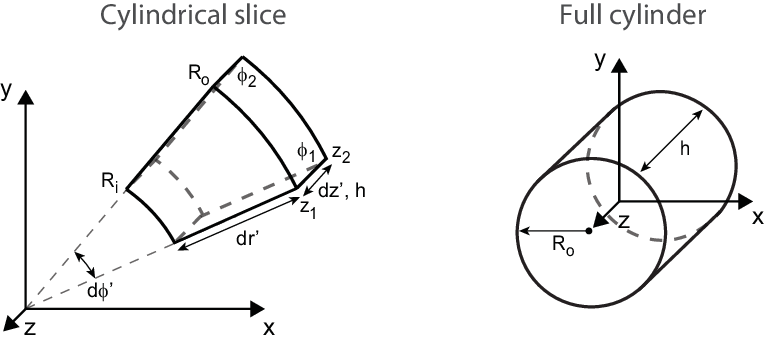}
   \caption{The cylinder slice and full cylinder considered. The origin is placed such that it is at the center of the cylinder, regardless of whether this is a slice or a full cylinder. The cylinder slice is defined by its extent in the radial, $R_i$ to $R_o$, and angular, $\phi_1$ to $\phi_2$, directions and by its length from $z_1=-h/2$ to $z_2=h/2$ where $h$ is the height of the slice. The full cylinder is defined by its radius, $R_o$ and length, $h$.}
   \label{Fig.Illustration}
\end{figure}

We consider a full cylinder or a slice of this, as illustrated in Fig. \ref{Fig.Illustration}. We have chosen a coordinate system identical to that of Refs. \cite{Nielsen_2020} and \cite{Slanovc_2022}. We consider in the following first the magnetic scalar potential from a uniformly magnetized cylindrical slice and after the special case of a uniformly magnetized full cylinder. The potential of a hollow cylinder is easily obtained as that of a full cylinder with the given magnetization minus that of an inner cylinder with the opposite magnetization, as potentials are additive.

The cylinder slice is placed such that the origin is at the center of the cylinder. The cylinder slice is defined by its extent in the radial, $R_i$ to $R_o$, and angular, $\phi_1$ to $\phi_2$, directions and by its length from $z_1=-h/2$ to $z_2=h/2$ where $h$ is the height of the slice. 

We will use a cylindrical coordinate system with $r$ denoting the radius, $\phi$ the angle, and $z$ the height. This also applies to the marked coordinates, which are the coordinates of a face that generates a field. To obtain the magnetic scalar potential from the cylindrical slice, the potentials generated by the six surfaces of the slice must be summed. There are three unique surfaces, one at constant $r'$, one at constant $\phi'$, and one at constant $z'$. As noted above, the magnetic surface charge is defined as $\sigma = \mathbf{\hat{n}} \cdot \mathbf{M}$, i.e., as the dot product of the surface normal vector with the magnetization vector. This quantity is homogeneous over each of the faces of the cylindrical slice.

The magnetization in Cartesian coordinates as well as expressed with spherical coordinates angles is given by \cite{Slanovc_2022}
\begin{eqnarray}
    \mathbf{M} &=& [M_x, M_y, M_z] \\ 
     &=& M\left(\textrm{cos}(\phi_M)\textrm{sin}(\theta_M)\mathbf{\hat{e}}_x+\textrm{sin}(\phi_M)\textrm{sin}(\theta_M)\mathbf{\hat{e}}_y\right. \nonumber\\ && \left.+\textrm{cos}(\theta_M)\mathbf{\hat{e}}_z\right) \nonumber
\end{eqnarray}

In cylindrical coordinates the distance is given by
\begin{equation}\label{Eq.cyl_dist}
    \|\mathbf{r}-\mathbf{r}'\| = \sqrt{r^2+r'^2-2rr'\textrm{cos}(\phi-\phi{}')+(z-z')^2}
\end{equation}
and the differential normal area element, $dS'$, is
\begin{equation}\label{Eq.dS}
    dS' = r'd\phi{}'dz'\mathbf{\hat{r'}} + dr'dz'\boldsymbol{\hat{\phi'}} + r'dr'd\phi'\mathbf{\hat{z'}}
\end{equation}
and the magnetic surface charge is \cite{Slanovc_2022}
\begin{eqnarray}\label{Eq.Surface_charge}
    \sigma_{r'} &=& \mp{}M\textrm{sin}(\theta_M)\textrm{cos}(\phi_M-\phi') \nonumber \\
    \sigma_{\phi'} &=& \mp{}M\textrm{sin}(\theta_M)\textrm{sin}(\phi_M-\phi') \nonumber \\
    \sigma_{z'} &=& \mp{}M\textrm{cos}(\theta_M)
\end{eqnarray}
where the $\mp{}$ is determined from the surface normal.

The analytical expressions for the magnetic scalar potential derived in the following can be quite complex, as is also the case for the magnetic field expression of a cylindrical slice \cite{Nielsen_2020,Slanovc_2022}. We introduce the following constants to ease the analytical expressions:
\begin{eqnarray}
a &=& -2 r r' \nonumber \\
b &=& r^2+r'^2+(z-z')^2 \nonumber \\
c &=& r^2 - 3r'^2 + 3b \nonumber \\
d &=& \sqrt{4r^2b - a^2} \nonumber \\
e &=& -2(r'^2 - b)^2 + \frac{3}{2}a^2 - 6br^2
\end{eqnarray}

Additionally, with $\mathrm{F}$ we denote the incomplete elliptic integral of the first kind, with $\mathrm{K}$ the complete elliptic integral of the first kind, with $\mathrm{E}$ the incomplete or complete elliptic integral of the second kind and with $\mathrm{\Pi}$ the incomplete or complete elliptic integral of the third kind. We define the following elliptic integrals
\begin{eqnarray}
    \mathbb{E} &=& \mathrm{E}\left(\frac{\phi-\phi{}'}{2}\middle|\frac{2\,a}{a+b}\right)  \nonumber\\
    \mathbb{F} &=& \mathrm{F}\left(\frac{\phi-\phi{}'}{2}\middle|\frac{2\,a}{a+b}\right) \nonumber \\
    \mathbb{P}1(\pm) &=& \Pi \left(1+\frac{{\left(z-z'\right)}^2}{{\pm{}d-2r^2-\left(z-z'\right)}^2};\frac{\phi-\phi{}'}{2}\middle|\frac{2\,a}{a+b}\right) \nonumber \\ 
    \mathbb{P}2(\pm) &=& \Pi \left(\frac{\left(a+b\right)\,{\left(z-z'\right)}^2}{\left(a-b\right)\,\left(\pm{}d- 2r^2 - (z-z')^2\right)};\right.  \nonumber\\
&&\left.-\mathrm{asin}\left(\sqrt{\frac{a-b}{a+b}}\mathrm{tan}\left(\frac{\phi-\phi{}'}{2}\right)\right)\middle|-\frac{a+b}{a-b}\right) \nonumber \\ 
    \mathbb{P}3 &=& \Pi \left(\frac{-2a}{(z-z')^2 - a - b}, \frac{\phi - \phi'}{2}, \frac{2a}{a + b}\right)
\end{eqnarray}

The $\pm$ indicates a possible sign change for the elliptic integral. In all terms, it is the factor $d$ that changes sign. We note that for the elliptic integrals of the third kind, $\mathbb{P}(n,\phi,m)$, the first argument in the integral might diverge. If this is the case, the value of the integral is simply $\mathbb{P}(n,\phi,m)\xrightarrow[n \to \infty]{}0$.

To express the magnetic scalar potential, we consider each of the three unique surfaces of the cylindrical slice in turn. The total potential is then the sum of these.

\subsection{The $z'=z_0$ surfaces}\label{Sec.z0slice}
We consider first the surface at constant $z'=z_0$, i.e. one of the horizontal surfaces of the slice. Without loss of generality, we choose the surface with the inwards-facing normal vector. Inserting Eqs. \eqref{Eq.cyl_dist} and \eqref{Eq.Surface_charge} into Eq. \eqref{eq:PhiMsolFlat} we get
\begin{eqnarray}\label{Eq.z_slice_integral}
    \Psi_{M,z0}(\bs{r})&=&-\frac{1}{4\pi} \int_{0}^{R}\int_{0}^{2\pi}\\
&&\frac{M\mathrm{cos}(\theta_M)r'}{\sqrt{r^2+r'^2-2rr'\mathrm{cos}(\phi-\phi{}')+(z-z')^2}}d\phi{}'dr'.\nonumber
\end{eqnarray}

The above integral is easily evaluated over $r'$, resulting in two terms. These can then, with some assistance from Rubi, the Rule-based Integrator \cite{Rich2018}, be integrated and the solution is given in terms of the previously defined elliptic integrals and a series of factors defined as
\begin{eqnarray}
    \mathcal{A}_{\pm} &=& -\frac{ \pm{}\sqrt{b\left(a^2 + d^2\right)} +2ra}{{\left(z-z'\right)}^2\sqrt{a+b}} \nonumber \\
    \mathcal{B}_{\pm} &=& \frac{r\sqrt{a+b}\,\left(\pm{}dc+2\left(\left(z-z'\right)^2+2\,r^2\right)^2+2r^2\left(z-z'\right)^2\right)}{\pm{}d\,\left(\pm{}d+2r^2\right)\,{\left(z-z'\right)}^2}  \nonumber \\
    \mathcal{C}_{\pm} &=& -\frac{r(\pm{}dc - e)\left( \pm{}d(b - a) - 2ar'^2 + 2br^2 - 4ar^2 + 2ab\right)}{\pm{}d\left(\pm{} d + 2r^2\right)^2(z-z')^2\sqrt{a + b}}  \nonumber \\
    \mathcal{D}_{\pm} &=& -\frac{2ar\left(\pm{}dc+e\right)\left(z-z'\right)^2}{\pm{}d\left(\pm{}d-2r^2\right)^2\left(\pm{}d-b+r'^2-r^2\right)\sqrt{a+b}}  \nonumber \\
    \mathcal{E} &=& -2i\frac{\left(ar'+2br\right)}{d\sqrt{b-a}}
\end{eqnarray}

As with the elliptic integrals, the $\pm$ indicates a change in sign.  

With the above defined elliptic integrals and factors, the integral in Eq. \eqref{Eq.z_slice_integral} at a given point $(r',\phi',z')$, which we term $q_{z_0}(r',\phi{}',z')$, is given as
\begin{eqnarray}
    q_{z_0}(r',\phi{}',z') &=& \frac{1}{4\pi}M\mathrm{cos}(\theta_M)r\Bigg(\\
&&\left.-\mathrm{atanh}\left(\frac{r'-r\,\cos\left(\phi -\phi'\right)}{\sqrt{b+a\,\cos\left(\phi -\phi'\right)}}\right)\nonumber\sin\left(\phi -\phi'\right) \right. \nonumber \\ 
&&    + \left(\mathcal{A}(+)+\mathcal{B}(+)+\mathcal{B}(-)-\frac{2\sqrt{a+b}}{r}\right)\mathbb{E} \nonumber\\
&& +\left(\mathcal{A}(-)+\mathcal{C}(+)+\mathcal{C}(-)\right)\mathbb{F} + \mathcal{D}(+)\mathbb{P}1(+) \nonumber \\
&&  +\mathcal{D}(-)\mathbb{P}1(-)+\mathcal{E}(\mathbb{P}2(+)-\mathbb{P}2(-))\Bigg)  \nonumber
    \label{Eq.q_for_zo}
\end{eqnarray}

We can reformulate this equation as a function that can be part of the demagnetization vector, which should only depends on one component of the magnetization, namely $M_z = M\textrm{cos}(\theta_M)$, as
\begin{eqnarray}
Q_{z_0}(r',\phi{}',z') &=& \frac{1}{4\pi}r\Bigg(-\mathrm{atanh}\left(\frac{r'-r\,\cos\left(\phi -\phi'\right)}{\sqrt{b+a\,\cos\left(\phi -\phi'\right)}}\right)\nonumber\\
  &&  \sin\left(\phi -\phi'\right) + \bigg(\mathcal{A}(+)+\mathcal{B}(+)+\mathcal{B}(-)\nonumber\\
 && \left.-\frac{2\sqrt{a+b}}{r}\right)\mathbb{E} +\big(\mathcal{A}(-)+\mathcal{C}(+) \nonumber \\
&&   +\mathcal{C}(-)\big)\mathbb{F} + \mathcal{D}(+)\mathbb{P}1(+)+\mathcal{D}(-)\mathbb{P}1(-) \nonumber \\
&& +\mathcal{E}(\mathbb{P}2(+)-\mathbb{P}2(-))\Bigg)\hat{\mathbf{e}}_z
\end{eqnarray}

The magnetic scalar potential from a $z'_0$ surface at height $z'=z_1=-h/2$ where $h$ is the height of the cylinder, that extends from an inner radius of $R_i$ to an outer radius of $R_o$, and with an angular extension from $\phi_1$ to $\phi_2$ is then given by
\begin{eqnarray}
\Psi_{M|z_1}  = -\left((q_{z_0}(R_o , \phi_2, -h/2) - q_{z_0}(R_o , \phi_1, -h/2)) \right. \nonumber \\
 \left. - (q_{z_0}(R_i, \phi_2, -h/2) - q_{z_0}(R_i, \phi_1, -h/2))\right)
\end{eqnarray}

Written as the demagnetization vector for this particular face, it is
\begin{eqnarray}
\mathbb{N}_{\Psi|z_1} &=& -\left((Q_{z_0}(R_o , \phi_2, -h/2) - Q_{z_0}(R_o , \phi_1, -h/2)) \right. \nonumber \\
&& \left. - (Q_{z_0}(R_i, \phi_2, -h/2) - Q_{z_0}(R_i, \phi_1, -h/2))\right) \nonumber \\
\Psi_{M|z_1}  &=& \mathbb{N}_{\Psi|z_1}\cdot{}\mathbf{M}
\end{eqnarray}

For the surface at height $z_2=+h/2$ we get the same expression, except there is an overall sign change due to the normal vector, i.e.
\begin{eqnarray}
\Psi_{M|z_2}  = +\left((q_{z_0}(R_o , \phi_2, h/2) - q_{z_0}(R_o , \phi_1, h/2)) \right. \nonumber \\
 \left. - (q_{z_0}(R_i, \phi_2, h/2) - q_{z_0}(R_i, \phi_1, h/2))\right)
\end{eqnarray}
or written as the demagnetization vector
\begin{eqnarray}
\mathbb{N}_{\Psi|z_2} &=& +\left((Q_{z_0}(R_o , \phi_2, -h/2) - Q_{z_0}(R_o , \phi_1, -h/2)) \right. \nonumber \\
&& \left. - (Q_{z_0}(R_i, \phi_2, -h/2) - Q_{z_0}(R_i, \phi_1, -h/2))\right) \nonumber\\
\Psi_{M|z_2}  &=& \mathbb{N}_{\Psi|z_2}\cdot{}\mathbf{M}
\end{eqnarray}

The above expressions must be considered individually in a number of special cases for which the functional expressions becomes singular. The special case of the full cylinder will be considered separately subsequently.

\subsubsection{Angles outside $\phi'=[\phi-\pi,\phi+\pi]$}\label{Sec.Misbehaving_angles}
The $\mathbb{P}2_{\pm}$ integral, unfortunately, contains a factor of $asin(tan)$ which incorrectly change sign compared to the true integral outside the range $\phi'=[\phi-\pi,\phi+\pi]$. Thus, if either $\phi_1$ or $\phi_2$ is outside this range, the function evaluation must be split to only consider this range. If this is the case we modify the angle as follows: if $\phi_1-\phi < -\pi$ then $\phi_{1,\n{new}} = ((\phi_1-\phi)-\pi)+(\phi-\pi)$ and $\phi_{2,\n{new}} = \phi_{2}$ while if $\phi_2-\phi > \pi$ then $\phi_{1,\n{new}} = \phi_{1}$ and $\phi_{2,\n{new}} = ((\phi_2-\phi)-\pi)+(\phi-\pi)$ and the analytical solution is now given by splitting the integral into two parts as follows for a surface at $z'=-h/2$:
\begin{eqnarray}
\mathbb{N}_{\Psi|z_1,\mathrm{part 1}} = -((Q_{z_0}(R_o , \phi+\pi  , -h/2) - Q_{z_0}(R_o , \phi_{1,\textrm{new}}, -h/2)) \nonumber \\
                      - (Q_{z_0}(R_i , \phi+\pi  , -h/2) - Q_{z_0}(R_i , \phi_{1,\textrm{new}}, -h/2))) \nonumber \\
\mathbb{N}_{\Psi|z_1,\mathrm{part 2}} = -((Q_{z_0}(R_o , \phi_{2,\textrm{new}}  , -h/2) - Q_{z_0}(R_o , \phi-\pi, -h/2)) \nonumber \\
                      - (Q_{z_0}(R_i , \phi_{2,\textrm{new}}  , -h/2) - Q_{z_0}(R_i , \phi-\pi, -h/2))) \nonumber                 
\end{eqnarray}
\vspace{-0.77cm}
\begin{eqnarray}
\Psi_{M|z_1} &=& (\mathbb{N}_{\Psi|z_1,\mathrm{part 1}} + \mathbb{N}_{\Psi|z_1,\mathrm{part 2}})\cdot{}\mathbf{M}
\end{eqnarray}
and similarly for $\Psi_{M|z_2}$.

However, as will be discussed in Sec. \ref{sec:FullCylinder} for the full cylinder, the value of the $\mathbb{P}2_{\pm}$ integral at the angular limits, $\phi'=\phi\pm{}\pi$, must anyway be determined. As will be shown subsequently in Eq. \eqref{Eq:FullCylinderIntegralsElliptics} this limit can be determined analytically. Knowing this limit, which we term $\mathbb{P}2_{c,\pm}$ we can account for the erroneous factor of the $\mathbb{P}2$ integral by subtracting this factor from the integral twice, i.e. for both the $\phi-\pi$ and $\phi+\pi$ limit. We thus get for the case of $\phi'<\phi-\pi$ or $\phi'>\phi+\pi$ that the correct value of the integral, $\mathbb{P}2_{\mathrm{correct,\pm}}$, is given by
\begin{eqnarray}
\mathbb{P}2_{\mathrm{correct,\pm}} = \mathbb{P}2_{\pm} - 2\mathbb{P}2_{c,\pm}
\end{eqnarray}

\subsubsection{The case of $r=0$}

When $r=0$ a number of the previously defined factors diverge or become undefined. However, the value of $Q_{z_0}(r',\phi{}',z')$ can be determined by noting that the prefractor of $r$ in Eq. \eqref{Eq.q_for_zo} ensures that all terms involving the $\Pi$-integrals go to zero. Left is the remaining terms, which for $r=0$ goes to 
\begin{eqnarray}\label{Eq:Limitr0forzsurfSlice}
    Q_{z_0}(r',\phi{}',z') = -\frac{1}{4\pi}(\phi-\phi')\sqrt{r'^2+(z-z')^2}\,\mathbf{\hat{z}}
\end{eqnarray}

\subsubsection{The case of $z-z'=0$}\label{Sec.casez0}
To find the value for $z-z'=0$ we can set $z-z'=0$ in the original integral in Eq. \eqref{Eq.z_slice_integral}, as we are not integrating with respect to this variable. The result is 
\begin{eqnarray}
    Q_{z_0}(r',\phi{}',z') = \frac{1}{4\pi}\Biggl(-|r - r'|\mathbb{E} + \textrm{sign}(r - r')(r + r')\mathbb{F} \nonumber \\
    \left.+ r\;\textrm{atanh}\left(\frac{-r' + r\,\textrm{cos}(\phi - \phi')}{\sqrt{b + a\,\textrm{cos}(\phi - \phi')}}\right)\textrm{sin}(\phi - \phi')\right)\hat{\mathbf{e}}_z
\end{eqnarray}

\subsubsection{The case of $r=0$ and $z-z'=0$}
From Eq. \eqref{Eq:Limitr0forzsurfSlice} it is easy to work out the limit in case of both $r=0$ and $z-z'=0$. Here we simply get
\begin{eqnarray}
    Q_{z_0}(r',\phi{}',z') = -\frac{1}{4\pi}(\phi-\phi')r'\mathbf{\hat{z}}
\end{eqnarray}

\subsection{The $r'=r_0$ surfaces}
For the surface at constant $r'=r_0$, the integration is with respect to $z'$ and $\phi'$. 
Using the expression for the normal vector to the $r'$-surface with the negative surface normal, the integral to be solved is
\begin{eqnarray}\label{Eq.Integral_r0_slice}
    \Psi_{M,r0}(\bs{r})&=&-\frac{1}{4\pi} \int\int\\&&\frac{M\mathrm{sin}(\theta_M)\mathrm{cos}(\phi_M-\phi')r'}{\sqrt{r^2+r'^2-2rr'\mathrm{cos}(\phi-\phi{}')+(z-z')^2}}d\phi{}'dz'.\nonumber
\end{eqnarray}

Performing the integral over $dz'$ is relatively easy, but the remaining integral over $d\phi'$ cannot be solved directly as it contains trigonometric functions that do not have the same arguments in terms of $\phi_M$, $\phi$ and $\phi'$. However, we can use a trigonometric identity to express the cosine as the sum of a cosine and sine, i.e. $\textrm{cos}(\phi_M-\phi') = \textrm{cos}(\phi_M)\textrm{cos}(\phi')+\textrm{sin}(\phi_M)\textrm{sin}(\phi')$.
If we first add and subtract $\phi$ in the identity we get $\textrm{cos}(\phi-\phi+\phi_M-\phi') = \textrm{cos}(\phi_M-\phi)\textrm{cos}(\phi-\phi')+\textrm{sin}(\phi_M-\phi)\textrm{sin}(\phi-\phi')$. This results in two integrals that can be solved. 

We define the following seven factors:
\begin{eqnarray}
    \mathcal{F} &=& -\frac{\left(z-z'\right)\sqrt{a+b}}{a} \nonumber \\
    \mathcal{G} &=& -\frac{{\left(z-z'\right)}\left(\left(z-z'\right)^2-2\,b\right)}{a\,\sqrt{a+b}} \nonumber \\
    \mathcal{H} &=& \frac{\left(z-z'\right)\left(\left(z-z'\right)^2+a-b\right)}{a\,\sqrt{a+b}} \nonumber \\
    \mathcal{I} &=& -\mathrm{atanh}\left(\frac{z-z'}{\sqrt{b+a\,\cos\left(\phi -\phi'\right)}}\right)\sin\left(\phi - \phi'\right)\nonumber \\
    \mathcal{J} &=& \frac{b-{\left(z-z'\right)}^2}{a}\mathrm{atanh}\left(\frac{\sqrt{b+a\,\cos\left(\phi -\phi'\right)}}{z-z'}\right) \nonumber \\
    \mathcal{K} &=& \mathrm{atanh}\left(\frac{z-z'}{\sqrt{b+a\,\cos\left(\phi - \phi'\right)}}\right)\cos\left(\phi - \phi'\right) \nonumber \\
    \mathcal{L} &=& \frac{\left(z-z'\right)\sqrt{b+a\,\cos\left(\phi - \phi'\right)}}{a}
\end{eqnarray}

With the previously defined elliptic integrals and the above factors, the complete integral at a given point $(r',\phi',z')$, which we term $q_{r_0}(r',\phi{}',z')$, is given as
\begin{eqnarray}
    q_{r_0}(r',\phi{}',z') &=& 
    -\frac{1}{4\pi}M\mathrm{sin}(\theta_M)r'\left(  \right. \nonumber \\
    &&\left.\mathrm{cos}(\phi_M-\phi)(\mathcal{F}\mathbb{E}+\mathcal{G}\mathbb{F}+\mathcal{H}\mathbb{P}3+\mathcal{I})\right. \nonumber \\
    && \left.- \mathrm{sin}(\phi_M-\phi)(\mathcal{J}+\mathcal{K}+\mathcal{L})\right)
    \label{Eq.q_for_ro}
\end{eqnarray}

Using trigonometric identities for $sin$ and $cos$, we can rewrite the above expression as a vector function, such that the scalar potential can again be expressed as the dot product between a demagnetization vector and the magnetization. The function is 
\begin{eqnarray}
    Q_{r_0}(r',\phi{}',z') &=& -\frac{1}{4\pi}r'\left(((\mathcal{F}\mathbb{E}+\mathcal{G}\mathbb{F}+\mathcal{H}\mathbb{P}3+\mathcal{I})\mathrm{cos}(\phi) \right. \nonumber \\
&& + (\mathcal{J}+\mathcal{K}+\mathcal{L})\mathrm{sin}(\phi))\hat{\mathbf{e}}_x \nonumber \\
    && +((\mathcal{F}\mathbb{E}+\mathcal{G}\mathbb{F}+\mathcal{H}\mathbb{P}3+\mathcal{I})\mathrm{sin}(\phi) \nonumber \\
&&\left. - (\mathcal{J}+\mathcal{K}+\mathcal{L})\mathrm{cos}(\phi))\hat{\mathbf{e}}_y\right)
\end{eqnarray}

The magnetic scalar potential from an $r'_0$-surface at radius $r'=R_i$ is
\begin{eqnarray}
\Psi_{M|R_i} = -\left((q_{r_0}(R_i , \phi_2, h/2) - q_{r_0}(R_i , \phi_1, h/2)) \right. \nonumber \\
 \left. - (q_{r_0}(R_i, \phi_2, -h/2) - q_{r_0}(R_i, \phi_1, -h/2))\right)
\end{eqnarray}

or written as the demagnetization vector for this particular face, it is
\begin{eqnarray}
\mathbb{N}_{\Psi|R_i} &=& -\left((Q_{r_0}(R_i , \phi_2, h/2) - Q_{r_0}(R_i , \phi_1, h/2)) \right. \nonumber \\
&& \left. - (Q_{r_0}(R_i, \phi_2, -h/2) - Q_{r_0}(R_i, \phi_1, -h/2))\right) \nonumber \\
\Psi_{M|R_i}  &=& \mathbb{N}_{\Psi|R_i}\cdot{}\mathbf{M}
\end{eqnarray}

For the surface at radius $r'=R_o$ where $R_o>R_i$ we get the same expression, except there is an overall sign change due to the normal vector, so we get
\begin{eqnarray}
\mathbb{N}_{\Psi|R_o} &=& +\left((Q_{r_0}(R_o , \phi_2, h/2) - Q_{r_0}(R_o , \phi_1, h/2)) \right. \nonumber \\
&& \left. - (Q_{r_0}(R_o, \phi_2, -h/2) - Q_{r_0}(R_o, \phi_1, -h/2))\right) \nonumber \\
\Psi_{M|R_o}  &=& \mathbb{N}_{\Psi|R_o}\cdot{}\mathbf{M}
\end{eqnarray}
As for the surface at constant $z'$, the above expressions must be considered individually in a number of special cases for which the functional expressions become singular, and the expressions must be determined in the limits. 

\subsubsection{The case of $r=0$}
One case is for $r=0$. Setting $r=0$ in Eq. \eqref{Eq.Integral_r0_slice} the integrals gives
\begin{eqnarray}
    Q_{r_0}(r',\phi{}',z') &=& -\frac{1}{4\pi}r'\mathrm{ln}\left(z -z' + \sqrt{r'^2 + (z - z')^2}\right)\nonumber\\
&&(\mathrm{sin}(\phi')\hat{\mathbf{e}}_x - \mathrm{cos}(\phi')\hat{\mathbf{e}}_y)\label{Eq.limitr0z}
\end{eqnarray}

\subsubsection{The case of $z-z'=0$}
Similar to Sec. \ref{Sec.casez0} to find the value for $z-z'=0$ we can set $z-z'=0$ in the integral in Eq. \eqref{Eq.Integral_r0_slice} and obtain 
\begin{eqnarray}
    Q_{r_0}(r',\phi{}',z') = -\frac{1}{4\pi}r'i\frac{b}{a}\frac{\pi}{2}(-\mathrm{sin}(\phi)\mathbf{\hat{x}} + \mathrm{cos}(\phi)\mathbf{\hat{y}})
\end{eqnarray}

\subsubsection{The case of $r=0$ and $z-z'=0$}
From the case of $r=0$, i.e. Eq. \eqref{Eq.limitr0z}, the case for both $r=0$ and $z-z'=0$ can easily be worked out to be
\begin{eqnarray}
    Q_{r_0}(r',\phi{}',z') = -\frac{1}{4\pi}r'\mathrm{ln}\left(r'\right)(\mathrm{sin}(\phi')\mathbf{\hat{x}} - \mathrm{cos}(\phi')\mathbf{\hat{y}})
\end{eqnarray}

\subsection{The $\phi'=\phi_0$ surfaces}
For the surface at constant $\phi'=\phi_0$, the integration is over $r'$ and $z'$. Note that these surfaces are only present for the cylindrical slice and not the full cylinder. The integral for the negative surface normal is
\begin{eqnarray}
    \Psi_{M,\phi0}(\bs{r})&=&-\frac{1}{4\pi} \int\int\\&&\frac{M\mathrm{sin}(\theta_M)\mathrm{sin}(\phi_M - \phi')}{\sqrt{r^2+r'^2-2rr'\mathrm{cos}(\phi-\phi{}')+(z-z')^2}}d\phi{}'dz'.\nonumber
\end{eqnarray}

Notice that because of the differential normal area element, Eq. \eqref{Eq.dS}, this integral does not have the $r'$-factor present for the $z'$ and $r'$ surfaces.

We define the following four factors:
\begin{eqnarray}
\mathcal{M} &=& ir\sin\left(\phi -\phi'\right)\nonumber\\
&&\mathrm{atanh}\left(\frac{\left(z-z'\right)\left(r'-r\,\cos\left(\phi -\phi'\right)\right)}{r\,\sin\left(\phi -\phi'\right)\sqrt{-b-a\,\cos\left(\phi -\phi'\right)}}\right) \nonumber \\
\mathcal{N} &=& \left(z-z'\right)\mathrm{atanh}\left(\frac{r\,\cos\left(\phi -\phi'\right)-r'}{\sqrt{b+a\,\cos\left(\phi -\phi'\right)}}\right)\nonumber \\
\mathcal{O} &=& -r'\,\mathrm{atanh}\left(\frac{z-z'}{\sqrt{b+a\,\cos\left(\phi -\phi'\right)}}\right)  \\
\mathcal{P} &=& r\,\cos\left(\phi -\phi'\right)\,\mathrm{atanh}\left(\frac{\sqrt{b+a\,\cos\left(\phi -\phi'\right)}}{z-z'}\right) \nonumber
\end{eqnarray}

The integral at a given point $(r',\phi',z')$, which we term $q_{\phi_0}(r',\phi{}',z')$, is then given by 
\begin{equation}
q_{\phi_0}(r',\phi{}',z')=\frac{1}{4\pi}M\mathrm{sin}(\theta_M)\mathrm{sin}(\phi_M - \phi')\left(\mathcal{M}+\mathcal{N}+\mathcal{O}+\mathcal{P}\right)
\end{equation}

Again we can rewrite the above equation to be able to express it in terms of the demagnetization vector as
\begin{equation}
Q_{\phi_0}(r',\phi{}',z')=-\frac{1}{4\pi}\left(\mathcal{M}+\mathcal{N}+\mathcal{O}+\mathcal{P}\right)(\mathrm{sin}(\phi')\hat{\mathbf{e}}_x - \mathrm{cos}(\phi')\hat{\mathbf{e}}_y)
\end{equation}

The magnetic scalar potential from a $\phi'_0$-surface at an angle $\phi'=\phi_1$ is given by
\begin{eqnarray}
\mathbb{N}_{\Psi|\phi_1}  &=& -\left((Q_{\phi_0}(R_o , \phi_1, h/2) - Q_{\phi_0}(R_o , \phi_1, -h/2)) \right. \nonumber \\
&& \left. - (Q_{\phi_0}(R_i, \phi_1, h/2) - Q_{\phi_0}(R_i, \phi_1, -h/2))\right) \nonumber \\
\Psi_{M|\phi_1}  &=& \mathbb{N}_{\Psi|\phi_1}\cdot{}\mathbf{M}
\end{eqnarray}

For the surface at angle $\phi'=\phi_2$ where $\phi_2>\phi_1$ we get the same expression, except there is an overall sign change due to the normal vector, so we get
\begin{eqnarray}
\mathbb{N}_{\Psi|\phi_2}  &=& +\left((Q_{\phi_0}(R_o , \phi_2, h/2) - Q_{\phi_0}(R_o , \phi_2, -h/2)) \right. \nonumber \\
&& \left. - (Q_{\phi_0}(R_i, \phi_2, h/2) - Q_{\phi_0}(R_i, \phi_2, -h/2))\right) \nonumber \\
\Psi_{M|\phi_2}  &=& \mathbb{N}_{\Psi|\phi_2}\cdot{}\mathbf{M}
\end{eqnarray}

\subsubsection{The case of $r=0$}
Similar to the cases for the other faces, the case of $r=0$ must be considered separately. We get
\begin{eqnarray}
Q_{\phi_0}(r',\phi{}',z') = -\frac{1}{4\pi}\left((z-z')\mathrm{atanh}\left(\frac{r'}{\sqrt{r'^2 + (z - z')^2}}\right)  \right. \nonumber \\
+ r'\mathrm{ln}\left(z - z' + \sqrt{r'^2 + (z - z')^2} \right)\Biggl)\nonumber \\
\left(-\mathrm{sin}(\phi')\hat{\mathbf{e}}_x + \mathrm{cos}(\phi')\hat{\mathbf{e}}_y\right)
\end{eqnarray}

\subsubsection{The case of $r=0$ and $z-z'=0$}
For the case of $r=0$ and $z-z'=0$ the above expression reduces to
\begin{eqnarray}
Q_{\phi_0}(r',\phi{}',z') = -\frac{1}{4\pi}r'\mathrm{ln}(r')\left(-\mathrm{sin}(\phi')\hat{\mathbf{e}}_x + \mathrm{cos}(\phi')\hat{\mathbf{e}}_y\right) 
\end{eqnarray}

\subsection{Validation}
The magnetic scalar potential from a cylindrical slice is given by the sum of the potential from its six surfaces,
\begin{equation}
\Psi_{M}  = \Psi_{M|R_i} + \Psi_{M|R_o} + \Psi_{M|z_1} + \Psi_{M|z_2} + \Psi_{M|\phi_1} + \Psi_{M|\phi_2}
\end{equation}

or
\begin{equation}
\Psi_{M}  = \left(\mathbb{N}_{\Psi|R_i} + \mathbb{N}_{\Psi|R_o} + \mathbb{N}_{\Psi|z_1} + \mathbb{N}_{\Psi|z_2} + \mathbb{N}_{\Psi|\phi_1} + \mathbb{N}_{\Psi|\phi_2}\right)\cdot{}\mathbf{M}\label{Eq.Final_slice}
\end{equation}

\begin{figure}[ht]
  \centering
  \includegraphics[width=1\linewidth]{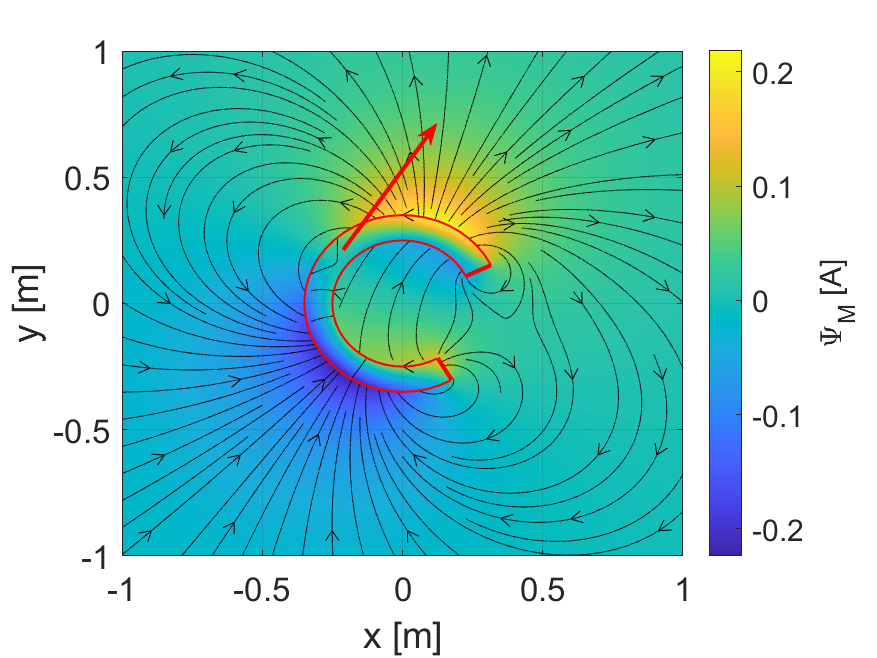}
    \caption{The magnetic scalar potential, $\Psi_M$, and the magnetic field lines generated by a cylindrical slice with an inner radius of $R_i = 0.25$ m, an outer radius of $R_o=0.35$ m, a height of $h= 0.7$ m and an angular extension from $\phi_1 = \pi/7$ to $\phi_2 = 2\pi-\pi/3$ and magnetization of $M = [2,\,  3,\, 4]$ A/m in the plane $z=0$ m. The magnetization direction is indicated by the red arrow, and the sides of the cylindrical slice are shown with the red lines.}
    \label{fig:Slice_4_0.001_z_0.0}
\end{figure}

To illustrate the solution, we consider a cylindrical slice defined by an inner radius of $R_i = 0.25$ m and an outer radius of $R_o=0.35$ m, a height of $h= 0.7$ m and an angular extension from $\phi_1 = \pi/7$ to $\phi_2 = 2\pi-\pi/3$, and with the slice centered on the origo. We consider a uniform magnetization of $M = [2,\,  3,\, 4]$ A/m. In Fig. \ref{fig:Slice_4_0.001_z_0.0} we show the magnetic scalar potential and the magnetic field lines computed from this in the slice $z=0$ m.

\begin{figure}[ht]
   \centering
  \includegraphics[width=1\linewidth]{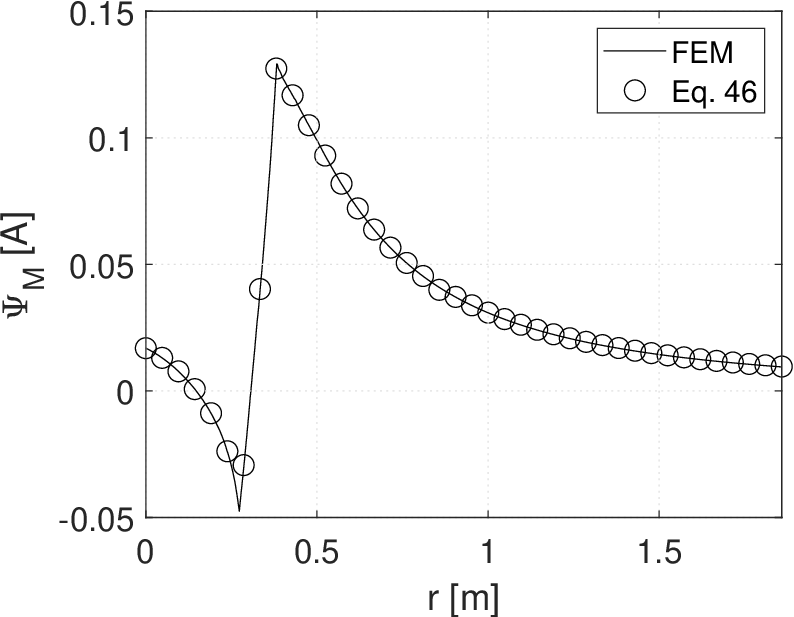}
   \caption{The magnetic scalar potential, $\Psi_M$, generated by a cylindrical slice with dimensions as given in Fig. \ref{fig:Slice_4_0.001_z_0.0} along the line $[0,\, 0,\, 0]\rightarrow[1.5,\, 0.8,\, 0.75]$ m as given by Eq. \eqref{Eq.Final_slice} and as computed using finite element modeling.}
   \label{fig:Potential_along_line_slice}
\end{figure}

To validate the analytical expression, the magnetic scalar potential has been computed using the finite element framework Comsol. In these simulations the number of mesh elements and the size of the computational domain have been chosen large enough that the results are converged, but a mesh/domain size convergence analysis is not presented here for brevity. We evaluate the scalar potential from the origin to a point located at $P = [x,\, y,\, z] = [1.5,\, 0.8,\, 0.75]$ m. In Fig. \ref{fig:Potential_along_line_slice} we show the magnetic scalar potential from the origin and to the point specified above. As can be seen there is a perfect agreement between the analytical expression and the finite element calculations, which also indicates that the latter is converged.

\section{The magnetic scalar potential of the full cylinder}\label{sec:FullCylinder}
To find the magnetic scalar potential for a full cylinder, we have to integrate over the full cylinder surface. This means that compared to the cylindrical slice the integral over the $\phi'$-surface disappear and there are three integrals left, one over the tube surface and two over the end surfaces, which will differ only in the value of $z'$. The cylinder is placed centered on the origin and is defined by its radius, $R_o$ and length, $h$ as shown in Fig. \ref{Fig.Illustration}.

The elliptic integrals defined for the cylinder slice simplifies for the full cylinder. These are given below. However, due to the periodicity of the elliptic integrals, as discussed in Sec. \ref{Sec.Misbehaving_angles}, the angular integrals must be performed from $\phi' = \phi-\pi$ to $\phi' = \phi+\pi$. For these values of $\phi'$, the elliptic integrals defined for the cylindrical slice reduce to the expressions below, with an additional factor of $\textrm{sign}(\phi-\phi{}')$ on all terms. However, when having taken the angular limits of $\phi' = \phi\pm\pi$ we see that all dependence on $\phi'$ is only in a factor in front of all the terms, the previously mentioned $\textrm{sign}(\phi-\phi{}')$ factor. When taking the integration limits over $\phi'$, which results in an additional minus sign, we are simply left with two times all the terms, as will be shown explicitly when the magnetic scalar potential is computed.

The elliptic integrals, denoted with a subscript ``c'', for the full cylinder are
\begin{eqnarray}\label{Eq:FullCylinderIntegralsElliptics}
    \mathbb{E}_c &=& \mathrm{E}\left(\frac{2\,a}{a+b}\right) \nonumber \\
    \mathbb{F}_c &=& \mathrm{K}\left(\frac{2\,a}{a+b}\right) \nonumber \\
    \mathbb{P}1_{c,\pm} &=& \Pi \left(1+\frac{{\left(z-z'\right)}^2}{{\pm{}d-2r^2-\left(z-z'\right)}^2}\middle|\frac{2\,a}{a+b}\right) \nonumber \\ 
    n_{\pm} &=& \frac{a + b}{a - b}\frac{(z-z')^2}{\pm{}d - 2r^2 - (z-z')^2} \nonumber \\
    \mathbb{P}2_{c,\pm} &=& 
    \frac{i}{1-n_\pm}\left(n_\pm\Pi \left(1-n_\pm,\frac{2a}{a - b}\right)-\mathrm{K}\left(\frac{2a}{a - b}\right)\right) \nonumber \\ 
    \mathbb{P}3_c &=& \Pi \left(\frac{-2a}{(z-z')^2 - a - b}, \frac{2a}{a + b}\right)
\end{eqnarray}

As for the elliptic integrals for the cylindrical slice, the first argument in the elliptic integrals, $\mathbb{P}(n,m)$, might diverge. In this case the same limit applies as for the slice, namely 
$\mathbb{P}(n,m)\xrightarrow[n \to \infty]{}0$.

Additionally, we must consider the case of $n=1$ for which the $\mathbb{P}2_c$ factor becomes singular. Taking the limit of $n\rightarrow{}1$ the result is 
\begin{equation}
    \mathbb{P}2_{c,\pm}(n=1) = 
    i\left(-\mathrm{K}\left(\frac{2a}{a - b}\right) + \frac{a - b}{2a}\left(\mathrm{K}\left(\frac{2a}{a - b}\right)-\mathrm{E}\left(\frac{2a}{a - b}\right)\right)\right)
\end{equation}

\subsection{The $z'=z_0$ surfaces}
For the surface at constant $z'=z_0$ for a full cylinder, we get a result identical to Sec. \ref{Sec.z0slice} except that the first term in Eq. \eqref{Eq.q_for_zo}, the $atanh$ term, is equal to zero. We thus get for the full cylinder

\begin{eqnarray}
    q_{c,z_0}(r',z') = && \frac{1}{4\pi}M\mathrm{cos}(\theta_M)r\Biggr( \nonumber \\
&&\left(\mathcal{A}(+)+\mathcal{B}(+)+\mathcal{B}(-)-\frac{2\sqrt{a+b}}{r}\right)\mathbb{E}_c \nonumber \\ 
&&     +\left(\mathcal{A}(-)+\mathcal{C}(+)+\mathcal{C}(-)\right)\mathbb{F}_c \nonumber \\ 
&& +\mathcal{D}(+)\mathbb{P}1_c(+)+\mathcal{D}(-)\mathbb{P}1_c(-)\nonumber \\ 
&&+\mathcal{E}(\mathbb{P}2_c(+)-\mathbb{P}2_c(-))\Biggr)  
\end{eqnarray}

Again, it is useful to rewrite this such that the magnetic scalar potential can be expressed as the dot product between a demagnetization vector and the magnetization. The necessary function is
\begin{eqnarray}
    Q_{c,z_0}(r',z') &=& \frac{1}{4\pi}r\Biggr(\biggr(\mathcal{A}(+)+\mathcal{B}(+)+\mathcal{B}(-) \nonumber\\
    &&-\frac{2\sqrt{a+b}}{r}\biggl)\mathbb{E}_c +\biggr(\mathcal{A}(-)+\mathcal{C}(+)  \nonumber \\ 
&&      +\mathcal{C}(-)\biggl)\mathbb{F}_c +\mathcal{D}(+)\mathbb{P}1_c(+) +\mathcal{D}(-)\mathbb{P}1_c(-)\nonumber \\ 
&& +\mathcal{E}(\mathbb{P}2_c(+)-\mathbb{P}2_c(-))\Biggr)\hat{\mathbf{e}}_z  \label{Eq.Qcz0}
\end{eqnarray}

The scalar potential from one $z'$-surface at height $z'=z_1=-h/2$ where $h$ is the height of the cylinder is then given by the expressions below, where the factor of two is a result of the previously mentioned discarded $\textrm{sign}(\phi-\phi{}')$ factor. The magnetic scalar potential is
\begin{equation}
\Psi_{M,c|z_1}  = -\left(-2q_{c,z_0}(R_o , -h/2) + 2q_{c,z_0}(0, -h/2)\right)
\end{equation}
or written as the demagnetization vector
\begin{eqnarray}
\mathbb{N}_{\Psi,c|z_1} &=& -\left(-2Q_{c,z_0}(R_o , -h/2) + 2Q_{c,z_0}(0, -h/2)\right) \nonumber \\
\Psi_{M,c|z_1}  &=& \mathbb{N}_{\Psi,c|z_1}\cdot{}\mathbf{M}
\end{eqnarray}

and similarly for height $z'=z_2=h/2$ with an opposite normal
written as the demagnetization vector
\begin{eqnarray}
\mathbb{N}_{\Psi,c|z_2} &=& +\left(-2Q_{c,z_0}(R_o , h/2) + 2Q_{c,z_0}(0, h/2)\right) \nonumber \\
\Psi_{M,c|z_2}  &=& \mathbb{N}_{\Psi,c|z_2}\cdot{}\mathbf{M}
\end{eqnarray}

\subsubsection{The case of $r'=0$}
One of the factors for a full cylinder is evaluating $Q_{c,z_0}$ for $r'=0$, i.e. the inner boundary of the integral. While this can be done using Eq. \eqref{Eq.Qcz0}, the expression simplifies significantly for $r'=0$ and is equal to
\begin{equation}
Q_{c,z_0}(0,z') = -\pi\left(\sqrt{b} + r\left(-\frac{\sqrt{n(+)}-1}{1-n(+)}+\frac{\sqrt{n(-)}-1}{1-n(-)}\right)\right)\hat{\mathbf{e}}_z
\end{equation}

\subsubsection{The case of $r=0$}
Similar to the cylindrical slice, we must consider the case of $r=0$. We get for the full cylinder 
\begin{eqnarray}
    Q_{c,z_0}(r',z') = -\frac{1}{4\pi}(\mathbb{E}_c + \mathbb{F}_c)\sqrt{r'^2+(z-z')^2}\mathbf{\hat{z}}
\end{eqnarray}

\subsubsection{The case of $z-z'=0$}
Similar to the cylindrical slice as in Sec. \ref{Sec.casez0}, we get for the full cylinder
\begin{eqnarray}
    Q_{c,z_0}(r',z') = \frac{1}{4\pi}\left(-|r - r'|\mathbb{E}_c + \textrm{sign}(r - r')(r + r')\mathbb{F}_c\right)\mathbf{\hat{z}}
\end{eqnarray}

\subsubsection{The case of $r=0$ and $z-z'=0$}
For the case of both $r=0$ and $z-z'=0$ we get
\begin{eqnarray}
    Q_{c,z_0}(r',z') = -\frac{1}{4\pi}\pi{}r'\mathbf{\hat{z}}
\end{eqnarray}

\subsection{The $r'_0$ surface}
Taking the integral limits for the full cylinder the first three factors compared to the cylindrical slice reduce as given below. The additional factors does not change sign when integration around from $\phi+\pi$ to $\phi-\pi$ and so they cancel out when taking the integration limits. 

Performing the integrals, the terms $\mathcal{I}$, $\mathcal{J}$, $\mathcal{K}$ and $\mathcal{L}$ are all equal to zero. Thus the complete integral at a given point $(r',\phi',z')$, which we term $q_{r_0}(r',\phi{}',z')$, is given as
\begin{eqnarray}
    q_{c,r_0}(r',z') &=& 
    -\frac{1}{4\pi}M\mathrm{sin}(\theta_M)r'
    \mathrm{cos}(\phi_M-\phi) \nonumber \\
&& (\mathcal{F}\mathbb{E}_c+\mathcal{G}\mathbb{F}_c+\mathcal{H}\mathbb{P}3_c)
    \label{Eq.q_for_c_ro}
\end{eqnarray}
or written as a vector to compute the demagnetization vector, 
\begin{eqnarray}
    Q_{c,r_0}(r',z') &=& -\frac{1}{4\pi}r'(\mathcal{F}\mathbb{E}_c+\mathcal{G}\mathbb{F}_c+\mathcal{H}\mathbb{P}3_c)\nonumber \\
&&\left(\mathrm{cos}(\phi)\hat{\mathbf{e}}_x+\mathrm{sin}(\phi)\hat{\mathbf{e}}_{y}\right)
\end{eqnarray}

The magnetic scalar potential from a $r'_0$ surface at radius $r'=R_o$ is
\begin{eqnarray}
\Psi_{M|R_o}  &=& -2q_{c,r_0}(R_o , h/2) + 2q_{c,r_0}(R_o , -h/2)
\end{eqnarray}
or written as the demagnetization vector
\begin{eqnarray}
\mathbb{N}_{\Psi,c|R_o} &=& -2Q_{c,r_0}(R_o , h/2) + 2Q_{c,r_0}(R_o , -h/2) \nonumber \\
\Psi_{M,c|R_o}  &=& \mathbb{N}_{\Psi,c|R_o}\cdot{}\mathbf{M}
\end{eqnarray}

\subsubsection{The case of $r=0$}
For the case of $r=0$, the potential is simply
\begin{eqnarray}
\Psi_{M,c|R_o}  &=& 0
\end{eqnarray}

\subsubsection{The case of $z-z'=0$}
For the case of $z-z'=0$, the potential is also 
\begin{eqnarray}
\Psi_{M,c|R_o}  &=& 0
\end{eqnarray}

\subsubsection{The case of $r=0$ and $z-z'=0$}
As a logical consequence of the two special cases above, the potential for $r=0$ and $z-z'=0$ is simply
\begin{eqnarray}
\Psi_{M,c|R_o}  &=& 0
\end{eqnarray}

\subsection{The $\phi'=\phi_0$ surfaces}
For a full cylinder, there are no surfaces with constant $\phi'=\phi_0$, and so this does not appear in the magnetic scalar potential.

\subsection{Validation}
The magnetic scalar potential from a full cylinder is given by
\begin{eqnarray}
\Psi_{M}  &=& \Psi_{M,c|R_o}  + \Psi_{M,c|z_1} + \Psi_{M,c|z_2} 
\end{eqnarray}
or 
\begin{eqnarray}
\Psi_{M}  &=& \left(\mathbb{N}_{\Psi,c|R_o} + \mathbb{N}_{\Psi,c|z_1} + \mathbb{N}_{\Psi,c|z_2}\right)\cdot{}\mathbf{M}\label{Eq.final_full}
\end{eqnarray}

\begin{figure}[ht]
    \centering
   \includegraphics[width=1\linewidth]{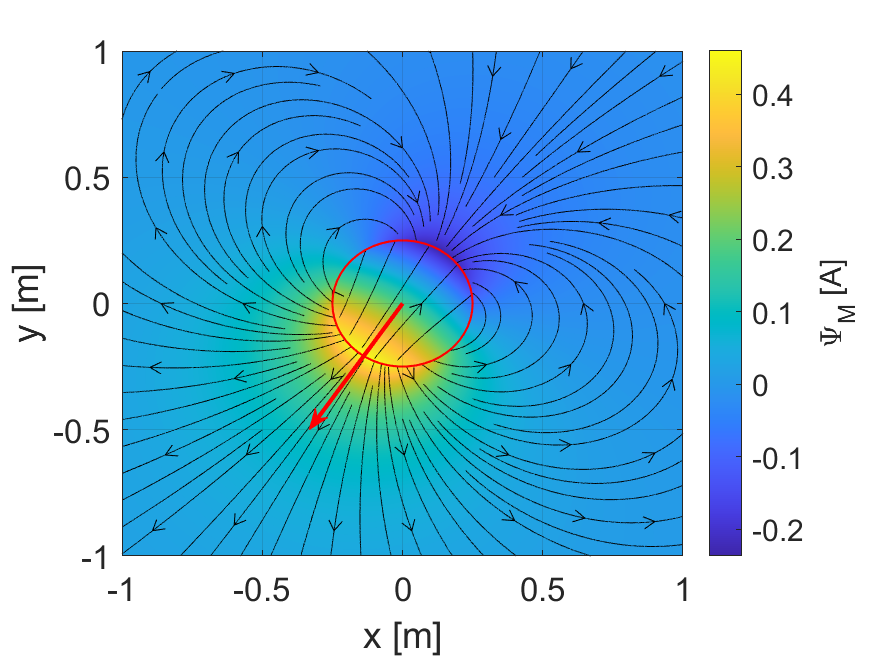}
   \caption{The magnetic scalar potential, $\Psi_M$, and the magnetic field lines generated by a cylinder with radius $R_o = 0.25$ m and height $h= 0.7$ m and magnetization of $M = [-2,\,  -3,\, 4]$ A/m in the plane $z=0.2$ m. The magnetization direction is indicated by the red arrow, and the side of the cylinder are shown with the red line.}
   \label{fig:Full_8_0.001_z_0.2}
\end{figure}

To illustrate the solution, we consider a full cylinder defined by a radius of $R_o = 0.25$ m and height of $h= 0.7$ m centered on the origo. We consider a magnetization of $M = [-2,\,  -3,\, 4]$ A/m. In Fig. \ref{fig:Full_8_0.001_z_0.2} we show the magnetic scalar potential and the magnetic field lines computed from this in the slice $z=0.2$ m.

\begin{figure}[ht]
  \centering
   \includegraphics[width=1\linewidth]{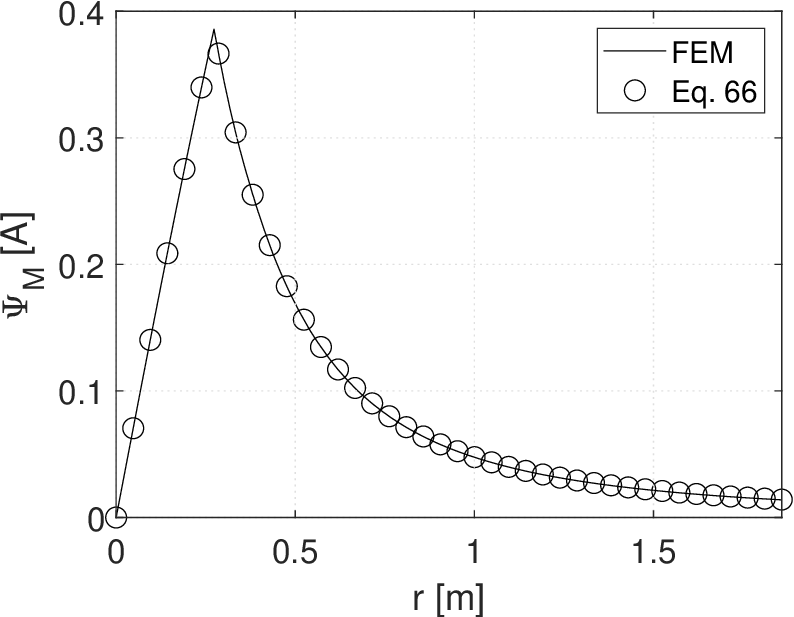}
   \caption{The magnetic scalar potential, $\Psi_M$, generated by a cylinder with radius $R_o = 0.25$ m and height $h= 0.7$ m and magnetization of $M = [-2,\,  -3,\, 4]$ A/m along the line $[0,\, 0,\, 0]\rightarrow[-1.5,\, -0.8,\, 0.75]$ m as given by Eq. \eqref{Eq.final_full} and as computed using finite element modeling.}
   \label{fig:Potential_along_line_full}
\end{figure}

Similar to the validation of the cylindrical slice, the magnetic scalar potential has been computed using the finite element framework Comsol. We evaluate the scalar potential from the origin to a point located at $P = [x,\, y,\, z] = $ \\ $ [-1.5,\, -0.8,\, 0.75]$ m. In Fig. \ref{fig:Potential_along_line_full} we show the magnetic scalar potential from the origin and to the point $P$ specified above. As was also the case for the cylindrical slice, a perfect agreement between the analytical expression and the finite element calculations can be seen.

\section*{Data statement}
The derived analytical expressions for the magnetic scalar potential for a cylindrical slice as well as for a full cylinder are available through the MagTense framework \cite{MagTense}, available at \href{https://www.magtense.org}{www.magtense.org} as are the examples shown in the figures in this work. Furthermore, the data shown in this work is also available at the data repository at Ref. \cite{Data_2025b}.

\section{Conclusion}
We have analytically solved Poisson's equation for the magnetic scalar potential generated by both a cylindrical slice as well as a full cylinder and determined a closed-form solution. The solution for both the cylindrical slice and the full cylinder are defined in terms of elliptic integrals. We show that the solution can be written as the dot product of a demagnetization vector, containing all the geometric information of the cylindrical slice or the full cylinder, and the magnetization. We validated the analytical expression by comparing with a finite element simulation and show that these agree perfectly.

\section*{Acknowledgements}
This work was supported by the Independent Research Fund Denmark, grant “Magnetic Enhancements through Nanoscale Orientation (METEOR)”, 1032-00251B, by the Villum Foundation Synergy project number 50091 entitled "Physics-aware machine learning" and by the Carlsberg Foundation Semper Ardens Advance project CF24-0920 entitled "Novel magnets through interdisiplinarity and nanocomposites".

\bibliographystyle{plain}

\end{document}